# Semantic Networks for Engineering Design: A Survey

A Preprint

**Ji Han**
School of Engineering
University of Liverpool
United Kingdom
ji.han@liverpool.ac.uk

**Serhad Sarica**
Data-Driven Innovation Lab
Singapore University of Technology and Design
Singapore, 487372
Serhad_sarica@mymail.sutd.edu.sg

**Feng Shi**
Amazon Web Services
United Kingdom
fenshi@amazon.co.uk

**Jianxi Luo**
Data-Driven Innovation Lab
Singapore University of Technology and Design
Singapore, 487372
luo@sutd.edu.sg

12$^{th}$ December, 2020

**Abstract**

There have been growing uses of semantic networks in the past decade, such as leveraging large-scale pre-trained graph knowledge databases for various natural language processing (NLP) tasks in engineering design research. Therefore, the paper provides a survey of the research that has employed semantic networks in the engineering design research community. The survey reveals that engineering design researchers have primarily relied on WordNet, ConceptNet, and other common-sense semantic network databases trained on non-engineering data sources to develop methods or tools for engineering design. Meanwhile, there are emerging efforts to mine large scale technical publication and patent databases to construct engineering-contextualized semantic network databases, e.g., B-Link and TechNet, to support NLP in engineering design. On this basis, we recommend future research directions for the construction and applications of engineering-related semantic networks in engineering design research and practice.

# 1 INTRODUCTION

Design knowledge retrieval, representation, and management are considered significant activities in engineering design because design is essentially a knowledge-intensive process (Bertola and Teixeira, 2003; Chandrasegaran et al., 2013). Digital knowledge bases have thereby been growingly employed to support engineering designers in the design process. These knowledge bases are often in the form of semantic networks. A semantic network is an artificial associative network representing knowledge in relation patterns of nodes and links interconnected in a graph structure (Sowa, 1992). Nodes in a semantic network represent specific knowledge pieces, concepts, or ideas (semantic entities), while the links refer to mental connections (semantic relations), which demonstrate how knowledge can be accessed from one another (Boden, 2004). It is shown that design information retrieval based on semantic networks outperforms the conventional keyword-based search (Li and Ramani, 2007).

In the engineering design literature, the most often used open-source, public semantic networks include WordNet (Miller, 1995), ConceptNet (Liu and Singh, 2004; Speer and Havasi, 2012; Speer et al., 2017), YAGO (Suchanek et al., 2007), and NELL (Never-Ending Language Learning) (Carlson et al., 2010; Mitchell et al., 2018). Such semantic networks serve as the knowledge base and digital infrastructure to support computational concept inferences for engineering knowledge discovery, learning, representation, synthesis, or evaluation. However, these semantic networks only involve general knowledge and relations, and were not designed specifically for engineering design. In recent years, there is an emerging interest in constructing new semantic networks based on engineering data sources and applying them as engineering knowledge bases for supporting engineering design knowledge discovery, analysis, representation, learning, and synthesis (Sarica et al., 2020).

The aim of this paper is to explore the current state of academic research and implementation of large-scale semantic networks as knowledge bases for engineering design, and illuminate potential directions for future research. To be more specific, the paper provides an overview of 1) the research that employs semantic networks for language processing in engineering design (what and how semantic networks are used as knowledge bases to provide computational design aids), and 2) the research that constructs engineering semantic networks (data sources and construction methods). On this basis, potential research directions for developing future engineering design semantic networks and their applications are proposed.

# 2 SEMANTIC NETWORKS AS KNOWLEDGE BASES FOR ENGINEERING DESIGN

Engineering design researchers extensively curated and/or utilized domain-specific and detailed knowledge bases which are not necessarily semantic networks. One important aspect of these knowledge bases is that they directly target a specific task or domain. For instance, Concept Generator (Bryant et al., 2006) is an automated design tool based on an algorithm using the Functional Basis (Otto and Wood, 1997; Hirtz et al., 2002) and employing an online design knowledge repository for producing feasible design concept variants. The design knowledge repository is not a semantic network and contains domain-specific knowledge only. Mukherjea et al. (2005) introduced the BioMedical Patent Semantic Web by annotating patents from the biomedical domain with entities from biomedical ontologies and retrieving relations between entities using a predefined set of patterns.

Some of these knowledge bases supported studies on Design-by-Analogy. Design-by-Analogy to Nature Engine (DANE) (Vattam et al., 2011; Goel et al., 2012) is a knowledge-based computational design tool supporting bio-inspired idea generation. The knowledge base used is a hand-built semantic network based on the SBF (Structure-Behaviour-Function) modelling framework, which contains a limited amount of domain-specific knowledge regarding biological and engineering systems. Analogy Finder, developed by McCaffrey and Spector (2017), could retrieve adaptable analogues from the US patent database for solving problems. The US patent database contains useful technical knowledge from patents but was not in the form of semantic networks. Idea Inspire 4.0 (Siddharth and Chakrabarti, 2018) is an idea generation support tool providing access to biological information in a human-curated knowledge base. The tool enhances its search capabilities for related words of the keyword provided by employing WordNet.

In addition, Georgiev et al. (2017) came up with a computational approach to produce ideas of new scenes by synthesizing existing scenes via thematic relations. A hand-built semantic network knowledge base containing thematic relations to store scenes was employed. Hu et al. (2017) developed an Intelligent Creative Conceptual Design System, which retrieves a domain-specific Function-Behaviour-Structure (FBS) knowledge cell library according to WordNet ontology. InnoGPS, developed by Luo et al. (2019), is a computer-aided design ideation support tool that provides rapid concept retrieval as inspirational stimuli and real-time evaluation of ideas generated. It uses a technology space map as the knowledge base, which is constructed based on patent data. He et al. (2019) constructed a semantic network of concepts based on their co-occurrences in a set of one thousand idea descriptions from an online crowdsourcing campaign via Mechanical Turk for reuse to inspire design ideation. concepTe (Acharya and Chakrabarti, 2020) is a decision-making support tool during the conceptual design stage offering aids in the designer's familiar domain, of which the knowledge base is grounded in the domain-agnostic SAPPhIRE model ontology.

In recent years, there are increasing applications of publicly available pre-trained large-scale semantic networks as the backend knowledge base for developing methods and tools for design ideation and analysis in the engineering design domain. WordNet (Miller et al., 1995) has been the most popular. For instance, WordTree (Linsey et al., 2012) uses brainstorming sessions and the WordNet's hierarchical structure to populate a tree structure, in which functional aspects of the design problem are represented with additional verbs to search for analogical solutions. Yoon et al. (2015) proposed a method to discover patents according to their function similarity assessed by leveraging WordNet's hierarchical structure. Cheong et al. (2017) extracted function knowledge from natural language texts utilizing WordNet and word2vec-based classification methods. Kan and Gero (2018) used WordNet for constructing linkographs to characterize innovative processes in design spaces. Georgiev and Georgiev (2018) developed WordNet-based metrics to measure divergence, polysemy, and creativity of new ideas. Goucher-Lambert and Cagan (2019) used semantic similarity and distance information in WordNet to categorize crowdsourced ideas as stimuli for design ideation. Nomaguchi et al. (2019) evaluated the novelty of function combinations in design ideas based on their semantic similarities in WordNet and a word2vec model trained on Wikipedia. A negative correlation between the human evaluations of novelty and the semantic similarity was reported. Liu et al. (2020) created a concept network by mining concepts from the technical documents related to a specific design problem and associating them via their world-embedding vectors and synset relations in the WordNet.

Other than WordNet, which was collectively built via human efforts, a few other free online knowledge bases have also been employed in design research and methodologies. For example, ConceptNet (Speer et al., 2017) is a large public knowledge graph automatically extracted from Wikipedia, built and maintained at MIT Media Lab. Yuan and Hsieh (2015) presented a tool using ConceptNet to support designers in framing the creation process for insight discovery. The Combinator, developed by Han et al. (2018a), is a creative idea generation support tool based on combinational creativity, which could produce combinational textual and pictorial stimuli. It involves a knowledge base constructed by extracting design keywords from design websites and associating them using the semantic relations in ConceptNet. The Retriever (Han et al., 2018b) employed ConceptNet as its sole knowledge base for supporting designers in creative idea generation via analogical reasoning. Han et al. (2020) also proposed to evaluate new ideas based on the semantic similarity of their elemental concepts using ConceptNet. Chen and Krishnamurthy (2020) proposed an interactive procedure to retrieve words and terms in ConceptNet to inspire designers. Camburn et al. (2019) proposed a set of new metrics for automatic evaluation of the natural language descriptions of a large number of crowdsourced design ideas, and their evaluation was based on the Freebase (Bollacker et al., 2008), another large knowledge database managed by Google.

These engineering design studies generally rely on common-sense knowledge bases, such as WordNet and ConceptNet, or language models not trained specifically for engineering. In fact, the engineers' perception of technical terms is biased and represented better by knowledge bases specifically trained on technological knowledge (Sarica et al., 2020). The growing uses of such public semantic network databases in the engineering design research and methodological developments have motivated the development of the semantic networks based on engineering data. For instance, Shi et al. (2017) mined and analysed nearly one million engineering papers in a span of 20 years from ScienceDirect to construct a large-scale semantic network, i.e., B-Link. Shi et al. (2017) and Chen et al. (2019) have utilized B-link to retrieve semantic level stimuli, synthesized together with images, to stimulate design ideation.

Sarica et al. (2020) constructed a technology semantic network (i.e., TechNet) consisting of more than 4 million technology-related terms that represent technical concepts in all domains of technology, and their semantic distance by exploiting the complete digitalized USPTO patent database from 1976 to 2017. The utilization of the complete patent database was aimed to ensure TechNet's comprehensiveness and the balanced coverage of knowledge in all domains of technology. In a benchmark comparison with other existing semantic network databases, including WordNet, ConceptNet, and B-Link, TechNet presented superior performances in term retrieval and inference tasks in the specific context of technology and engineering (Sarica et al., 2020). TechNet has been utilized to augment patent search (Sarica et al., 2019a), technology forecasting (Sarica et al., 2019b), and idea evaluation (Han et al., 2020).

Table 1 summarizes the engineering design studies using semantic networks, highlighting the purpose of the method or tool, the employed semantic network, and the type of knowledge contained.

Table 1. Semantic networks employed as knowledge bases in engineering design research

| | Purpose | Semantic Network | Knowledge Type |
|---|---|---|---|
| DANE (Vattam et al., 2011; Goel et al., 2012) | Idea generation | Hand-built based on SBF (Structure-Behaviour-Function) modelling framework | Domain-specific knowledge |
| (Yuan and Hsieh, 2015) | Idea generation | ConceptNet | General knowledge |
| (Georgiev et al., 2017) | Idea generation | Hand-built extracting thematic relations | Domain-specific knowledge |
| ICCDS (Hu et al., 2017) | Idea generation | WordNet FBS (Function-Behaviour-Structure) knowledge cell library | General knowledge from WordNet, domain-specific knowledge from the FBS knowledge cell library |
| (Cheong et al., 2017) | Knowledge extraction | WordNet and word2vec | General knowledge |
| Combinator (Han et al., 2018a) | Idea generation | ConceptNet | General knowledge from ConceptNet, domain-specific knowledge from the Combinator database |
| Idea Inspire 4.0 (Siddharth and Chakrabarti, 2018) | Idea generation | WordNet | General knowledge from WordNet, domain-specific knowledge from the Idea Inspire 4.0 database |
| Retriever (Han et al., 2018b) | Idea generation | ConceptNet | General knowledge |
| (Chen et al., 2019) | Idea generation | B-Link | Technical knowledge from academic papers |
| InnoGPS (Luo et al., 2019) | Idea generation and evaluation | Technology space map | Technical knowledge from the total patent database |
| concepTe (Acharya and Chakrabarti, 2020) | Decision-making at the conceptual design stage | SAPPhIRE model ontology | Domain-agnostic knowledge |
| (Chen and Krishnamurthy, 2020) | Idea generation | ConceptNet | General knowledge |
| TechNet (Sarica et al., 2019a; Sarica et al., 2019b; Han et al., 2020) | Idea generation, evaluation, prior art search | TechNet | Technical knowledge from the total patent database |

## 3 CONSTRUCTION OF SEMANTIC NETWORKS

The semantic networks that have been employed in engineering design research were constructed using different statistical approaches (e.g., hand-built, supervised, unsupervised) and based on different data sources (e.g., Wikipedia, Google News, Elsevier publication data, USPTO patent database). Table 2 presents a summary. In general, the construction of semantic networks requires the extraction of the entities from the raw data sources and statistically establishing the semantic relations among entities.

*Table 2. The data sources and construction methods of primary semantic networks*

|  | Construction Approach | Data Source | Relations | Engineering Related |
|---|---|---|---|---|
| WordNet (Miller, 1995) | Hand-build |  | Synonymy, hyponymy, meronymy, troponymy, antonymy | No |
| ConceptNet (Liu and Singh, 2004; Speer and Havasi, 2012; Speer et al., 2017) | Unsupervised | Open Mind Common Sense, DBPedia, Wiktionary, Open Multilingual WordNet, OpenCyc, GWAP Project | 34 types of relations: e.g. RelatedTo, FormOf, IsA, PartOf, HasA, UsedFor | No |
| YAGO (Suchanek et al., 2007) | Partial hand-build and unsupervised | Wikipedia, WordNet | 76 predefined relations | No |
| NELL (Carlson et al., 2010; Mitchell et al., 2018) | Semi-supervised | Web content | 461 different types of relations | No |
| Knowledge Vault (Dong et al., 2014) | Supervised | Web content | 4469 different types of relations | No |
| Pre-trained word2vec (Mikolov et al., 2013) | Unsupervised | Google News | Cosine similarity | No |
| Pre-trained GloVe (Pennington et al., 2014) | Unsupervised | Wikipedia, Gigaword, Common Crawl | Cosine similarity | No |
| B-Link (Shi et al., 2017) | Unsupervised | Academic papers, design blogs | Normalized network distance | Yes |
| TechNet (Sarica et al., 2020) | Unsupervised | Patents | Cosine similarity | Yes |

WordNet (Miller, 1995) is a large-scale lexical database of English constructed by experts through manually retrieving sets of cognitive synonyms and relations such as synonymy, hyponymy, and meronymy. ConceptNet (Liu and Singh, 2004; Speer and Havasi, 2012; Speer et al., 2017) is a knowledge graph built via unsupervised learning. It connects words and phrases retrieved from common-sense resources, including WordNet, Wikipedia, Wiktionary, and games with a purpose, via common-sense relations, e.g., PartOf, UsedFor, and IsA. YAGO (Suchanek et al., 2007) contains general knowledge automatically retrieved from Wikipedia and WordNet to fit a set of manually defined relations. NELL (Carlson et al., 2010; Mitchell et al., 2018) employs an infinite loop analogous to an Expectation-Maximization algorithm for semi-supervised learning of information in web pages. Knowledge Vault (Dong et al., 2014) uses supervised learning to fit probabilistic binary classifiers for fusing distinct data retrieved from web contents. Word2vec (Mikolov et al., 2013) is a popular pre-trained word embedding vector database, using a neural network for deriving the vector representations of words from Google News. GloVe (Pennington et al., 2014) is another popular pre-trained word embedding database that derives relations based on global statistics of co-occurrence counts of words from Wikipedia, Gigaword, and Common Crawl. B-Link (Shi et al., 2017) was developed using unsupervised learning by applying probability and velocity network analysis to correlate concepts retrieved from academic papers and design blogs. TechNet (Sarica et al., 2020) was derived using NLP techniques to extract terms from massive technical patent texts, as well as recent word embedding algorithms (i.e. word2vec and GloVe) to vectorise the terms and establish the semantic relations in the vector space.

Among these primary semantic networks, only WordNet was created by using a hand-built approach. The construction of hand-built semantic networks, which are often domain-specific and contain a limited amount of knowledge, is usually labour-intensive and time-consuming (Ahmed et al., 2006; Li et al., 2008), such as the SBF-based semantic network used in DANE (Vattam et al., 2011; Goel et al., 2012) and the thematic semantic network used by Georgiev et al. (2017). Knowledge Vault was developed using a supervised and semi-automatic approach that often requires human efforts. Supervised models need to be trained manually on large scale corpora before they could automatically recognise semantic entities and extract semantic relations. However, these models could only recognise the types of relations predefined in the training sets, making supervised learning challenging to construct semantic networks for engineering design that contains diverse engineering relations. By contrast, YAGO, pre-trained word2vec, pre-trained GloVe, B-Link, and TechNet were all constructed using unsupervised approaches to extract semantic relations from texts automatically.

As shown in Table 2, only B-Link and TechNet were trained using engineering related data sources, i.e., academic papers and patents, while the others employed general or common-sense knowledge data sources, e.g., Wikipedia and Google News. In addition to B-Link and TechNet, Li et al. (2005) proposed a partial-unsupervised approach for generating an engineering design domain-specific ontology, which uses basic NLP techniques and semantic analysis to retrieve knowledge from design documents and map them to a pre-structured ontology model. Li et al. (2009) came up with a partial-unsupervised approach to develop engineering ontologies assisted with a semi-automatic acquisition tool, using pre-processed engineering documents, i.e., catalogue descriptions, technical reports, and engineers' notebooks. Lim et al. (2010, 2011), and Liu et al. (2013) presented an unsupervised faceted information search and retrieval framework for creating product family ontology. Glier et al. (2014) developed an unsupervised method to identify text passages for designers by employing a text mining algorithm trained using survey data. Munoz and Tucker (2016) created an unsupervised semantic network of lecture content by indicating the relation between two words based on their sequential appearance within a given context window. However, these studies were not aimed at constructing large-scale comprehensive semantic networks for engineering design and setting up to serve as infrastructure to support prospective engineering design studies actively.

## 4  PROPOSITIONS FOR FUTURE RESEARCH DIRECTIONS

The most widely employed knowledge bases in engineering design literature are the common-sense semantic networks and lexical databases, such as WordNet and ConceptNet, which have been trained on non-engineering data sources. These common-sense semantic networks do not contain the necessary engineering design knowledge with contextual relations to support engineering design and analysis. Meanwhile, there are emerging efforts in constructing large-scale comprehensive engineering-contextualised semantic networks for engineering design applications by training the networks on technical publication (Shi et al., 2017) and patent databases (Sarica et al., 2020), which contain engineering design knowledge. These semantic networks have been used for supporting idea generation and evaluation, design information retrieval, augmenting prior art search, and technology forecasting. They can be potentially used as infrastructures to support an extensive range of engineering design applications.

The rapid advancements in NLP may provide new and better means to mine engineering data and learn engineering knowledge for constructing semantic networks in the context of engineering design. Recently, there is a surge of language models that uses deep neural network architectures, unlike word2vec and GloVe, to produce unfixed but context-aware word embeddings, such as ELMo (Peters et al., 2018) by AlienNLP, ULMFiT (Howard and Ruder, 2018) by fast.ai, Generative Pretrained Transformer (GPT, GPT-2, GPT-3) (Radford et al., 2018a; Radford et al., 2018b) by OpenAI, and BERT (Vaswani et al., 2017), XLNet (Yang et al., 2019), ALBERT (Lan et al., 2019) by Google. These models are pre-trained on very large corpora, letting researchers and practitioners fine-tune them with considerably small datasets to achieve downstream tasks, such as domain-specific text classification, named-entity detection, and sentiment detection. These models have resulted in record-breaking performances in various common NLP tasks and can be adopted to enhance the semantic networks and NLP tasks in the context of engineering design.

A critical limitation of large-scale and comprehensive semantic networks (e.g., B-Link and TechNet) is that the relations are one-dimensional. The entities are interconnected with weighted links indicating their semantic similarities. In contrast, domain-specific ontological databases allow drawing specialized and domain-specific qualitative relations among entities (Gero and Kannengiesser, 2014), while they lack generalizability. Knowledge graphs generally pose a trade-off between coverage and specificity (Zaveri et al., 2016) and aim to create a model of the real world by covering knowledge from a wide variety of fields, with continuous expansions of online data and constructions of relatively generalizable links between the entities stored (Paulheim, 2016). These advantages of knowledge graphs provide relational information that could be understood easily by both computers and humans. Besides, supported by language models, the structure of knowledge graphs informs AI tasks, such as knowledge search and discovery, summarization, reasoning, and question answering. Google, Facebook, IBM, e-bay, Netflix, Amazon, and many other companies alike have all developed knowledge graphs to power their machine learning and artificial intelligence engines. Likewise, comprehensive knowledge graphs trained on engineering design data are also expected to inform and augment engineering design.

To summarize, we recommend three future research directions of semantic networks for advancing technical language processing in engineering design, which involves:

- **Research Direction 1**: To extend the use of comprehensive large-scale semantic networks of technological knowledge, such as B-Link (Shi et al., 2017) and TechNet (Sarica et al., 2020), in engineering design.
- **Research Direction 2**: To apply up-to-date data science and NLP techniques, such as transformer-based language modelling architectures (e.g., ELMO, BERT, and GPT) to better capture semantic relations in the context of engineering design.
- **Research Direction 3**: To develop a comprehensive knowledge graph based on engineering knowledge data, which can evolve naturally, by constructing necessary pipelines for managing continuous information flow.

# 5 CONCLUDING REMARKS

This study contributes to the growing literature on data-driven and NLP-based engineering design analytics. In particular, we advocate using semantic networks trained on engineering data, in contrast to the common-sense semantic networks, for engineering design research and applications, and point to strategic directions for future developments of technology semantic networks. The public pre-trained large-scale technology semantic networks, e.g. TechNet and B-Link, may serve as an infrastructure for a wide range of artificial intelligence applications related to technology and engineering.

**REFERENCES**

Acharya, S. and Chakrabarti, A. (2020), "A conceptual tool for environmentally benign design: development and evaluation of a "proof of concept"", *Artificial Intelligence for Engineering Design, Analysis and Manufacturing*, Vol. 34 No.1, pp. 30-44. http://dx.doi.org/10.1017/S0890060419000313

Ahmed, S., Kim, S. and Wallace, K.M. (2006), "A Methodology for Creating Ontologies for Engineering Design", *Journal of Computing and Information Science in Engineering*, Vol. 7 No. 2, pp. 132-140. http://dx.doi.org/10.1115/1.2720879

Bertola, P. and Teixeira, J.C. (2003), "Design as a knowledge agent: How design as a knowledge process is embedded into organizations to foster innovation", *Design Studies*, Vol. 24 No. 2, pp. 181-194. https://doi.org/10.1016/S0142-694X(02)00036-4

Boden, M.A. (2004), *The creative mind: Myths and mechanisms*, 2 ed., London, UK: Routledge.

Bollacker, K., Evans, C., Paritosh, P., Sturge, T., and Taylor, J. (2008), "Freebase: A Collaboratively Created Graph Database for Structuring Human Knowledge", *Proceedings of the 2008 ACM SIGMOD International Conference on Management of Data, AcM*, pp. 1247–1250

Bryant, C.R., McAdams, D.A., Stone, R.B., Kurtoglu, T. and Campbell, M.I. (2006), "A Validation Study of an Automated Concept Generator Design Tool", *ASME 2006 International Design Engineering Technical Conferences and Computers and Information in Engineering Conference*, pp. 283-294. http://dx.doi.org/10.1115/DETC2006-99489

Camburn, B., He, Y., Raviselvam, S., Luo, J. and Wood, K. (2019), "Evaluating Crowdsourced Design Concepts With Machine Learning", *ASME 2019 International Design Engineering Technical Conferences and Computers and Information in Engineering Conference*. http://dx.doi.org/10.1115/detc2019-97285

Carlson, A., Betteridge, J., Kisiel, B., Settles, B., Hruschka, E.R. and Mitchell, T.M. (2010), "Toward an architecture for never-ending language learning", *Proceedings of the Twenty-Fourth AAAI Conference on Artificial Intelligence*, Atlanta, Georgia, AAAI Press, pp. 1306–1313.

Chandrasegaran, S.K., Ramani, K., Sriram, R.D., Horváth, I., Bernard, A., Harik, R.F. and Gao, W. (2013), "The evolution, challenges, and future of knowledge representation in product design systems", *Computer-Aided Design*, Vol. 45 No.2, pp. 204-228. http://dx.doi.org/10.1016/j.cad.2012.08.006

Chen, L., Wang, P., Dong, H., Shi, F., Han, J., Guo, Y., Childs, P.R.N., Xiao, J. and Wu, C. (2019), "An artificial intelligence based data-driven approach for design ideation", *Journal of Visual Communication and Image Representation*, Vol. 61, pp. 10-22. https://doi.org/10.1016/j.jvcir.2019.02.009

Chen, T.-J. and Krishnamurthy, V.R. (2020), "Investigating a Mixed-Initiative Workflow for Digital Mind-Mapping", *Journal of Mechanical Design*, Vol. 142 No.10. http://dx.doi.org/10.1115/1.4046808

Cheong, H., Li, W., Cheung, A., Nogueira, A. and Iorio, F. (2017), "Automated Extraction of Function Knowledge From Text", *Journal of Mechanical Design*, Vol. 139 No. 11, p. 111407. http://dx.doi.org/10.1115/1.4037817

Dong, X., Gabrilovich, E., Heitz, G., Horn, W., Lao, N., Murphy, K., Strohmann, T., Sun, S. and Zhang, W. (2014), "Knowledge vault: a web-scale approach to probabilistic knowledge fusion", *Proceedings of the 20th ACM SIGKDD International Conference on Knowledge Discovery and Data Mining*, New York, New York, USA, Association for Computing Machinery, pp. 601–610. http://dx.doi.org/10.1145/2623330.2623623

Georgiev, G.V. and Georgiev, D.D. (2018), "Enhancing user creativity: Semantic measures for idea generation", *Knowledge-Based Systems*, Vol. 151, pp. 1-15. https://doi.org/10.1016/j.knosys.2018.03.016

Georgiev, G.V., Sumitani, N. and Taura, T. (2017), "Methodology for creating new scenes through the use of thematic relations for innovative designs", *International Journal of Design Creativity and Innovation*, Vol. 5 No.1-2, pp. 78-94. http://dx.doi.org/10.1080/21650349.2015.1119658

Gero, J.S. and Kannengiesser, U. (2014), "The Function-Behaviour-Structure Ontology of Design", *An Anthology of Theories and Models of Design*, Springer London, London, pp. 263–283.

Glier, M.W., McAdams, D.A. and Linsey, J.S. (2014), "Exploring Automated Text Classification to Improve Keyword Corpus Search Results for Bioinspired Design", *Journal of Mechanical Design*, Vol. 136 No. 11. http://dx.doi.org/10.1115/1.4028167

Goel, A.K., Vattam, S., Wiltgen, B. and Helms, M. (2012), "Cognitive, collaborative, conceptual and creative — Four characteristics of the next generation of knowledge-based CAD systems: A study in biologically inspired design", *Computer-Aided Design*, Vol. 44 No. 10, pp. 879-900. https://doi.org/10.1016/j.cad.2011.03.010

Goucher-Lambert, K. and Cagan, J. (2019), "Crowdsourcing inspiration: Using crowd generated inspirational stimuli to support designer ideation", *Design Studies*, Vol. 61, pp. 1-29. https://doi.org/10.1016/j.destud.2019.01.001

Han, J., Forbes, H., Shi, F., Hao, J. and Schaefer, D. (2020), "A data-driven approach for creative concept generation and evaluation", *Proceedings of the Design Society: DESIGN Conference*. Cambridge University Press, Vol. 1, pp. 167–176. https://doi.org/10.1017/dsd.2020.5

Han, J., Shi, F., Chen, L. and Childs, P.R.N. (2018a), "The Combinator – a computer-based tool for creative idea generation based on a simulation approach", *Design Science*, Vol. 4, p. e11. http://dx.doi.org/10.1017/dsj.2018.7

Han, J., Shi, F., Chen, L. and Childs, P.R.N. (2018b), "A computational tool for creative idea generation based on analogical reasoning and ontology", *Artificial Intelligence for Engineering Design, Analysis and Manufacturing*, Vol. 32 No. 4, pp. 462-477. http://dx.doi.org/10.1017/S0890060418000082

He, Y., Camburn, B., Liu, H., Luo, J., Yang, M., and Wood, K. (2019), "Mining and Representing the Concept Space of Existing Ideas for Directed Ideation", *Journal of Mechanical Design,* Vol. 141 No.12, p. 121101. https://doi.org/10.1115/1.4044399

Hirtz, J., Stone, R.B., McAdams, D.A., Szykman, S. and Wood, K.L. (2002), "A functional basis for engineering design: Reconciling and evolving previous efforts", *Research in Engineering Design*, Vol. 13 No. 2, pp. 65-82. http://dx.doi.org/10.1007/s00163-001-0008-3

Howard, J. and Ruder, S. (2018), "Universal Language Model Fine-tuning for Text Classification", arXiv preprint arXiv:1801.06146

Hu, J., Ma, J., Feng, J.-F. and Peng, Y.-H. (2017), "Research on new creative conceptual design system using adapted case-based reasoning technique", *Artificial Intelligence for Engineering Design, Analysis and Manufacturing*, Vol. 31 No.1, pp. 16-29. http://dx.doi.org/10.1017/S0890060416000159

Kan, J.W.T. and Gero, J.S. (2018), "Characterizing innovative processes in design spaces through measuring the information entropy of empirical data from protocol studies", *Artificial Intelligence for Engineering Design, Analysis and Manufacturing*, Vol. 32 No. 1, pp. 32-43. http://dx.doi.org/10.1017/S0890060416000548

Lan, Z., Chen, M., Goodman, S., Gimpel, K., Sharma, P. and Soricut, R. (2019), "ALBERT: A Lite BERT for Self-supervised Learning of Language Representations", arXiv preprint arXiv:1909.11942v6

Li, Z. and Ramani, K. (2007), "Ontology-based design information extraction and retrieval", *Artificial Intelligence for Engineering Design, Analysis and Manufacturing*, Vol. 21 No. 2, pp. 137-154. http://dx.doi.org/10.1017/S0890060407070199

Li, Z., Liu, M., Anderson, D.C. and Ramani, K. (2005), "Semantics-Based Design Knowledge Annotation and Retrieval", *ASME 2005 International Design Engineering Technical Conferences and Computers and Information in Engineering Conference*, pp. 799-808. http://dx.doi.org/10.1115/detc2005-85107

Li, Z., Raskin, V. and Ramani, K. (2008), "Developing Engineering Ontology for Information Retrieval", *Journal of Computing and Information Science in Engineering*, Vol. 8 No. 1, p. 011003. http://dx.doi.org/10.1115/1.2830851

Li, Z., Yang, M.C. and Ramani, K. (2009), "A methodology for engineering ontology acquisition and validation", *Artificial Intelligence for Engineering Design, Analysis and Manufacturing*, Vol. 23 No. 1, pp. 37-51. http://dx.doi.org/10.1017/S0890060409000092

Lim, S.C.J., Liu, Y. and Lee, W.B. (2010), "Multi-facet product information search and retrieval using semantically annotated product family ontology", *Information Processing & Management*, Vol. 46 No. 4, pp. 479-493. https://doi.org/10.1016/j.ipm.2009.09.001

Lim, S.C.J., Liu, Y. and Lee, W.B. (2011), "A methodology for building a semantically annotated multi-faceted ontology for product family modelling", *Advanced Engineering Informatics*, Vol. 25 No. 2, pp. 147-161. https://doi.org/10.1016/j.aei.2010.07.005

Linsey, J. S., Markman, A. B., and Wood, K. L. (2012), "Design by Analogy: A Study of the WordTree Method for Problem Re-Representation", *Journal of Mechanical Design*, Vol. 134 No. 4, p. 041009. https://doi.org/10.1115/1.4006145

Liu, H. and Singh, P. (2004), "ConceptNet — A practical Common-sense reasoning tool-kit", *BT Technology Journal*, Vol. 22 No. 4, pp. 211-226. http://dx.doi.org/10.1023/b:bttj.0000047600.45421.6d

Liu, Q., Wang, K., Li, Y., and Liu, Y. (2020), "Data-Driven Concept Network for Inspiring Designers' Idea Generation", *Journal of Computing and Information Science in Engineering*, 20(3): 031004. https://doi.org/10.1115/1.4046207

Liu, Y., Lim, S.C.J. and Lee, W.B. (2013), "Product Family Design Through Ontology-Based Faceted Component Analysis, Selection, and Optimization", *Journal of Mechanical Design*, Vol. 135 No. 8, p. 081007. https://doi.org/10.1115/1.4023632

Luo, J., Sarica, S. and Wood, K.L. (2019), "Computer-Aided Design Ideation Using InnoGPS", *Proceedings of the ASME 2019 International Design Engineering Technical Conferences and Computers and Information in Engineering Conference*, p. V02AT03A011. http://doi.org/10.1115/detc2019-97587

McCaffrey, T. and Spector, L. (2017), "An approach to human–machine collaboration in innovation", *Artificial Intelligence for Engineering Design, Analysis and Manufacturing*, Vol. 32 No. 1, pp. 1-15. http://dx.doi.org/10.1017/S0890060416000524

Mikolov, T., Chen, K., Corrado, G. and Dean, J. (2013), "Efficient estimation of word representations in vector space", arXiv preprint arXiv:1301.3781

Miller, G.A. (1995), "WordNet: a lexical database for English", *Communications of the ACM*, Vol. 38 No. 11, pp. 39–41. http://dx.doi.org/10.1145/219717.219748

Mitchell, T., Cohen, W., Hruschka, E., Talukdar, P., Yang, B., Betteridge, J., Carlson, A., Dalvi, B., Gardner, M., Kisiel, B., Krishnamurthy, J., Lao, N., Mazaitis, K., Mohamed, T., Nakashole, N., Platanios, E., Ritter, A., Samadi, M., Settles, B., Wang, R., Wijaya, D., Gupta, A., Chen, X., Saparov, A., Greaves, M. and Welling, J. (2018), "Never-ending learning", *Communications of the ACM*, Vol. 61 No.5, pp. 103–115. http://dx.doi.org/10.1145/3191513.

Mukherjea, S., Bamba B. and Kankar P. (2005), "Information retrieval and knowledge discovery utilizing a biomedical patent semantic Web", *IEEE Transactions on Knowledge and Data Engineering*, Vol. 17 No. 8, pp. 1099-1110. https://doi.org/10.1109/TKDE.2005.130

Munoz, D. and Tucker, C.S. (2016), "Modeling the Semantic Structure of Textually Derived Learning Content and its Impact on Recipients' Response States", *Journal of Mechanical Design*, Vol. 138 No. 4. http://dx.doi.org/10.1115/1.4032398

Nomaguchi, Y., Kawahara, T., Shoda, K. and Fujita, K. (2019), "Assessing Concept Novelty Potential with Lexical and Distributional Word Similarity for Innovative Design", *Proceedings of the Design Society: International Conference on Engineering Design*, Vol. 1 No. 1, pp. 1413–1422. doi: 10.1017/dsi.2019.147

Otto, K. and Wood, K. (1997), "Conceptual and configuration design of products and assemblies", *ASM Handbook, Materials Selection and Design*, Vol. 20, pp. 15-32.

Paulheim, H. (2016), "Knowledge graph refinement: A survey of approaches and evaluation methods", *Semantic Web*, Vol. 8 No. 3, pp. 489–508.

Pennington, J., Socher, R. and Manning, C.D. (2014), "Glove: Global vectors for word representation", in *Proceedings of the 2014 conference on empirical methods in natural language processing (EMNLP)*, pp. 1532-1543.

Peters, M.E., Neumann, M., Iyyer, M., Gardner, M., Clark, C., Lee, K. and Zettlemoyer, L. (2018), "Deep contextualized word representations", *NAACL HLT 2018 - 2018 Conference of the North American Chapter of the Association for Computational Linguistics: Human Language Technologies - Proceedings of the Conference*, Vol. 1, pp. 2227–2237.

Radford, A., Narasimhan, K., Salimans, T. and Sutskever, I. (2018a), "Improving Language Understanding by Generative Pre-Training", OpenAI, available at: https://gluebenchmark.com/leaderboard%0Ahttps://s3-us-west-2.amazonaws.com/openai-assets/research-covers/language-unsupervised/language_understanding_paper.pdf

Radford, A., Wu, J., Child, R., Luan, D., Amodei, D. and Sutskever, I. (2018b), "Language Models Are Unsupervised Multitask Learners", OpenAI.

Sarica S., Song B., Low E., and Luo J. (2019a), "Engineering Knowledge Graph for Keyword Discovery in Patent Search", *Proceedings of the Design Society: International Conference on Engineering Design*, Vol. 1 No. 1, pp. 2249–2258. https://doi.org/10.1017/dsi.2019.231

Sarica S., Song B., Luo J., and Wood K. (2019b), "Technology Knowledge Graph for Design Exploration: Application to Designing the Future of Flying Cars", *Proceedings of the ASME 2019 International Design Engineering Technical Conferences and Computers and Information in Engineering Conference*, p. V001T02A028. https://doi.org/10.1115/DETC2019-97605

Sarica, S., Luo, J. and Wood, K.L. (2020), "TechNet: Technology semantic network based on patent data", *Expert Systems with Applications*, Vol. 142, p. 112995. https://doi.org/10.1016/j.eswa.2019.112995.

Shi, F., Chen, L., Han, J. and Childs, P. (2017), "A Data-Driven Text Mining and Semantic Network Analysis for Design Information Retrieval", *Journal of Mechanical Design*, Vol. 139 No. 11, p. 111402. http://dx.doi.org/10.1115/1.4037649

Siddharth, L. and Chakrabarti, A. (2018), "Evaluating the impact of Idea-Inspire 4.0 on analogical transfer of concepts", *Artificial Intelligence for Engineering Design, Analysis and Manufacturing*, Vol. 32 No. 4, pp. 431-448. http://dx.doi.org/10.1017/S0890060418000136

Sowa, J.F. (1992), "Semantic networks" in Shapiro, S. C., ed., *Encyclopaedia of artificial intelligence*, 2 ed., New York: John Wiley & Sons, pp. 1493–1511.

Speer, R. and Havasi, C. (2012), "Representing general relational knowledge in ConceptNet 5", *Proceedings of the Eight International Conference on Language Resources and Evaluation*.

Speer, R., Chin, J. and Havasi, C. (2017), "ConceptNet 5.5: an open multilingual graph of general knowledge", *Proceedings of the Thirty-First AAAI Conference on Artificial Intelligence*, San Francisco, California, USA, 3298212: AAAI Press, pp. 4444-4451.

Suchanek, F.M., Kasneci, G. and Weikum, G. (2007), "Yago: a core of semantic knowledge", *Proceedings of the 16th international conference on World Wide Web*, Banff, Alberta, Canada, Association for Computing Machinery, pp. 697–706. http://dx.doi.org/10.1145/1242572.1242667

Vaswani, A., Shazeer, N., Parmar, N., Uszkoreit, J., Jones, L., Gomez, A.N., Kaiser, Ł., et al. (2017), "Attention is all you need", *Advances in Neural Information Processing Systems*, pp. 5999–6009.

Vattam, S., Wiltgen, B., Helms, M., Goel, A.K. and Yen, J. (2011), "DANE: Fostering Creativity in and through Biologically Inspired Design", *Proceeding of the First International Conference on Design Creativity*, pp. 115-122.

Yang, Z., Dai, Z., Yang, Y. and Carbonell, J. (2019), "XLNet : Generalized Autoregressive Pretraining for Language Understanding", arXiv preprint arXiv:1906.08237

Yoon, J., Park, H., Seo, W., Lee, J.-M., Coh, B.-y. and Kim, J. (2015), "Technology opportunity discovery (TOD) from existing technologies and products: A function-based TOD framework", *Technological Forecasting and Social Change*, Vol. 100, pp. 153-167. https://doi.org/10.1016/j.techfore.2015.04.012.

Yuan, S.-T.D. and Hsieh, P.-K. (2015), "Using association reasoning tool to achieve semantic reframing of service design insight discovery", *Design Studies*, Vol. 40, pp. 143-175. https://doi.org/10.1016/j.destud.2015.07.001

Zaveri, A., Rula, A., Maurino, A., Pietrobon, R., Lehmann, J. and Auer, S. (2016), "Quality assessment for Linked Data: A Survey", *Semantic Web*, Vol. 7 No. 1, pp. 63-93.